\documentclass[12pt,preprint]{aastex}

\shorttitle{GRB Jet from NDAF}
\shortauthors{Kawanaka, Piran \& Krolik}

\usepackage{epsf}
\usepackage{epsfig}
\usepackage{amsmath}

\begin{document}

\title{Jet Luminosity From Neutrino-Dominated Accretion Flows in Gamma-Ray Bursts}
\author{Norita Kawanaka\altaffilmark{1a}, Tsvi Piran\altaffilmark{1b} \& Julian H. Krolik\altaffilmark{2c}}
\altaffiltext{1}{Racah Institute for Physics, The Hebrew University, Jerusalem, 91904, Israel}
\altaffiltext{2}{Physics and Astronomy Department, Johns Hopkins University, Baltimore, MD 21218, USA} 
\email{(a) norita@phys.huji.ac.il; (b) tsvi.piran@mail.huji.ac.il; (c) jhk@jhu.edu}

\begin{abstract}
A hyperaccretion disk formed around a stellar mass black hole is a plausible model for the central engine that powers gamma-ray bursts (GRBs).  If the central black hole rotates and a poloidal magnetic field threads its horizon, a powerful relativistic jet may be driven by a process resembling the Blandford-Znajek mechanism.  We estimate the luminosity of such a jet assuming that the poloidal magnetic field strength is comparable to the inner accretion disk pressure.  We show that the jet efficiency attains its maximal value when the accretion flow is cooled via optically-thin neutrino emission.  The jet luminosity is much larger than the energy deposition through neutrino-antineutrino annihilation ($\nu\bar{\nu}\rightarrow e^+e^-$) provided that the black hole is spinning rapidly enough.  When the accretion rate onto a rapidly spinning black hole { is large enough ($\gtrsim 0.003-0.01M_{\odot}{\rm s}^{-1}$), the predicted jet luminosity is sufficient to drive a GRB.}
\end{abstract}


\section{Introduction}
Gamma-ray bursts (GRBs) are the most luminous objects in the Universe, releasing $\gtrsim 10^{51}{\rm erg}$ in a few seconds.  The spectral features and lightcurves of their prompt and afterglow emission imply that they are produced in ultrarelativistic jets.  The central engines most likely to launch these jets are hyperaccreting black holes \citep{narayan+92, narayan+01}.  In this model, a relativistic jet emerges from a system with a massive accretion disk ($0.1-1M_{\odot}$) surrounding a stellar-mass black hole, a system expected to form after a cataclysmic event such as the gravitational collapse of a massive star or a merger of a neutron star-neutron star binary \citep{piran99}.

The observed luminosity requires a very large mass accretion rate, $\sim 0.01-10M_{\odot}{\rm s}^{-1}$.  In such a case the accretion flow is extremely optically thick with respect to photons and it cannot cool efficiently via radiation.  However, since the density and temperature are very large ($\rho \gtrsim 10^7{\rm g}~{\rm cm}^{-3}$, $T \gtrsim 10^{10}{\rm K}$) when the mass accretion rate is sufficiently high, neutrino cooling becomes efficient.  Such disks are `neutrino-dominated accretion flows' (NDAFs; \citealp{popham+99, narayan+01, dimatteo+02, kohrimineshige02, kohri+05, gu+06, chenbeloborodov07, kawanakamineshige07, liu+07, kawanakakohri12}; see also Chapter 10.6 of \citealp{kato+08}).

The accreting system must launch a relativistic jet to produce a GRB.  Neutrino pair annihilation ($\nu \bar{\nu} \rightarrow e^+e^-$; \citealp{eichler+89, ruffert+97, ruffertjanka98, popham+99, asanofukuyama00, asanofukuyama01, miller+03, birkl+07, harikae+09}) and magnetohydrodynamical mechanism such as Blandford-Znajek process \citep{blandfordznajek77, mckinneygammie04, hawleykrolik06, tchekhovskoy+08, tchekhovskoy+11, nagataki09} are the two processes most discussed in the literature.  The goal of this work is to estimate the luminosity of the jets emerging from a hyperaccretion flow driven by an MHD mechanism, as well as the luminosity of thermal neutrinos emitted from the disk.  The first is estimated using the maximal magnetic field sustainable on the event horizon, which is expected to be limited by the pressure in the disk near the ISCO, the innermost stable circular orbit \citep{krolikpiran11, krolikpiran12}.  The second we calculate directly from the density, pressure, and lepton number of the material in the disk.

In Section~2 we predict the jet power in the MHD model by solving for the structure of a surrounding NDAF and compare this luminosity to the energy deposition rate via neutrino pair annihilation above the accretion flow.  We discuss the parameter dependence of our model and its application to GRB jets in Section~3.  We summarize our results in Section~4. 

\section{Model}

We consider a rotating black hole whose horizon is threaded by a large-scale poloidal magnetic field \citep{blandfordznajek77}.  The Poynting luminosity expected from such a system is $\sim c(B^2/8\pi)R_g^2$, where $B$ is the poloidal magnetic field strength on the horizon and $R_g=GM_{\rm BH}/c^2$ is the gravitational radius of the central black hole \citep{mckinney06, hawleykrolik06}.  Following \cite{krolikpiran11} we estimate the jet luminosity as
\begin{eqnarray}
L_{\rm jet}=f(a/M_{\rm BH})c(B^2/8\pi)R_g^2, \label{jetlum}
\end{eqnarray}
where $f(a/M_{\rm BH})$ is an increasing function of $|a/M_{\rm BH}|$ whose exact form depends on the specific field configuration (e.g. \citealp{blandfordznajek77, tchekhovskoy+08, tchekhovskoy+11}).
There have been some attempts to calculate $f(a/M_{\rm BH})$ analytically for specified field geometries \citep{blandfordznajek77, tchekhovskoy+10}, but these configurations were not required to be consistent with the dynamics of the surrounding accretion flow.  They were also relatively simple poloidal topologies, but as \cite{beckwith+08} showed, high jet power requires that the vertical component of the magnetic field maintain the same sign for long times; in other words, low-order poloidal topologies are the most favorable to strong jets.  Consequently, it is possible for $f(a/M_{\rm BH})$ to vary over an extremely wide range, from very small numbers to $\sim O(1)$.  When the topology is optimal and $a/M \gtrsim 0.9$, the $f(a/M_{\rm BH})$ arising from a dynamically self-consistent field structure can be $\gtrsim 0.05$, and even larger for spin parameters closer to unity \citep{hawleykrolik06}.

\cite{beckwith+09} argued that the magnetic pressure near the horizon may be limited by the inner disk pressure.  The magnetic field energy may therefore be estimated (or at least bounded) by the disk pressure $p_{\rm disk}$ near the ISCO:
\begin{eqnarray}
\beta_h \frac{B^2}{8\pi}=p_{\rm disk}(R_{\rm in}),
\end{eqnarray}
where $R_{\rm in}$
is the radius of the pressure reference point
and $\beta_h$ is the ratio of the midplane pressure at $R_{\rm in}$ to the magnetic pressure in the black hole's stretched horizon.

To evaluate the midplane pressure of the hyperaccretion flow, we solve for the disk structure in the innermost region ($R\sim O(R_g)$), where the jet is expected to be launched.  As we show below, the behavior of an hyperaccretion disk is mainly determined by its mass accretion rate $\dot{M}$.  To highlight what we believe to be the principal physical mechanisms (and also eliminate mathematical complications), we adopt a highly simplified model for this structure in the expectation that the trends we uncover while exploring a factor of $10^5$ range in $\dot M$ will be robust with respect to refinements in the model.  In particular, we assume that both the disk dynamics and the gravitational potential are Newtonian despite the fact that this near a black hole general relativistic effects are important.  We also assume that the system is stationary and axisymmetric.  We adopt the Shakura-Sunyaev formalism for the disk structure, which formally applies only to thin disks, assumes that the inflow speed is very small compared to the orbital speed, and parameterizes the $r$-$\phi$ component of the stress \citep{shakurasunyaev73}.  { When the disk becomes geometrically thick, in the advection-dominated regimes, we use in our analytic estimates (but not in the numerical solution) a dimensional analysis.  Remarkably the results are similar up to a numerical coefficient to those obtained using the Shakura \& Sunyaev model.}  We further set the correction factor $X$ appearing in Equation~\ref{angcons} to always be unity, when it should in fact be somewhat less than that.  These are rough approximations, but we nonetheless expect our results to give the right qualitative picture in this scenario.

Here we present the basic disk equations for the density $\rho$, the temperature $T$, and the scale height $H$.  These are the expressions for mass conservation, angular momentum conservation, energy conservation, and hydrostatic balance applied at $R_{\rm in}$:
\begin{eqnarray}
\dot{M}&=&-2\pi R_{\rm in} \Sigma v_R, \label{masscons} \\
2\alpha H p_{\rm disk} 
&=&\frac{\dot{M}\Omega(R_{\rm in})}{2\pi}X(R/R_g), \label{angcons} \\
Q^+&=&Q^-, \label{energy} \\
\frac{p_{\rm disk}}{\rho}&=&\Omega(R_{\rm in})^2 H^2, \label{hydrostat} \\
\end{eqnarray}
where $\Sigma$, $v_R$, $\Omega(R)$, and $\alpha$ denote the surface density ($=2\rho H$), radial velocity, angular velocity ($=GM_{\rm BH}/R^3$), and ratio of integrated stress to integrated pressure, respectively.  In Newtonian dynamics, the factor $X(R/R_g)$ accounts for the reduction in stress due to the net angular momentum flux through the disk; although the net angular momentum flux per unit mass accreted has traditionally been identified with the specific angular momentum of orbits at the ISCO \cite{nt73}, it also depends upon MHD effects \citep{krolik99,gammie99,krolik+05}.

The pressure $p_{\rm disk}$ is the sum of the contributions from radiation, baryonic gas, degenerate electrons, and neutrinos (if they are trapped):
\begin{eqnarray}
p_{\rm disk}&\simeq &\frac{11}{12}aT^4+\frac{\rho k_B T}{m_p}+\frac{2\pi h c}{3}\left( \frac{3}{8\pi m_p} \right)^{4/3}\left( Y_e \rho \right)^{4/3} \nonumber \\
&&+\frac{u_{\nu}}{3},
\end{eqnarray}
where $a\simeq 7.56\times 10^{-15}{\rm erg}~{\rm cm}^{-3}~{\rm deg}^{-4}$ is the radiation constant, $k_B$ is the Boltzman constant, and $m_p$ is the mass of a proton.  The first term on the right hand side represents the contribution from both photons and relativistic electron-positron pairs; the coefficient is 11/12 when the temperature is higher than $\sim 10^{10}{\rm K}$.  In the third term, which corresponds to degeneracy pressure, $Y_e$ is the electron fraction, which should be evaluated from the $\beta$-equilibrium condition\footnote{Note that this approximation is valid only when the flow is optically thick with respect of $\nu_e$, and when neutrino emission is not efficient the electron fraction should be nearly equal to 0.5.  In the numerical calculations we make use of a bridging formula in evaluating $Y_e$ to connect these two extreme regimes smoothly.}
\begin{eqnarray}
\frac{1-Y_e}{Y_e}=\exp \left(-\frac{Q-\mu_e}{k_B T} \right),
\end{eqnarray}
where $Q\simeq 1.29{\rm MeV}$ is the difference between the neutron and proton masses, and $\mu_e$ is the chemical potential of the electrons (we assume that the chemical potential of neutrinos can be ignored and that heavy nuclei are totally dissociated in the innermost region of the accretion flow; see \citealp{kawanakamineshige07}).  The last term represents trapped neutrino pressure; $u_{\nu}$ is the neutrino energy density, defined below.

$Q^+$ is the heating rate per unit area:
\begin{eqnarray}
Q^+&\simeq &\frac{3GM_{\rm BH}\dot{M}}{4\pi R^3} \nonumber \\
&\simeq & 9.8\times 10^{36}{\rm erg}~{\rm cm}^{-2}~{\rm s}^{-1}~(\dot{M}/10^{-3}M_{\odot}{\rm s}^{-1}) \nonumber \\
&&\times (M_{\rm BH}/3M_{\odot})^{-2}(R/6R_g)^{-3}. \label{energy2}
\end{eqnarray}
$Q^-$ is the cooling rate per unit area, including both neutrino and advective cooling: $Q^-=Q^-_{\nu}+Q^-_{\rm adv}$.  The advective cooling rate $Q^-_{\rm adv}$, which equals the inward flux of the gas energy along the flow, can be described as:
\begin{eqnarray}
Q^-_{\rm adv}\equiv T\Sigma v_R\frac{ds}{dR}\simeq T\Sigma v_R\frac{s}{R_{\rm in}}.
\end{eqnarray}
Here $s$ denotes the entropy per unit mass.  There are several significant neutrino emission processes expected in such a hyperaccretion flow (see also \citealp{kawanakamineshige07} and references therein): electron/positron capture on nuclei ($p+e^- \rightarrow n+\nu_e$; $n+e^+ \rightarrow p+\bar{\nu}_e$, also known as the Urca process), electron-positron pair annihilation ($e^- + e^+ \rightarrow \nu_i + \bar{\nu}_i$, where $i$ represents electron-, mu-, and tau-type neutrinos), nucleon-nucleon bremsstrahlung ($n+n\rightarrow n+n+\nu_i+\bar{\nu}_i$), and plasmon decay ($\tilde{\gamma}\rightarrow \nu_e+\bar{\nu}_e$).  We should also take into account the neutrino optical depth in the accretion flow, which can be larger than unity when the mass accretion rate is sufficiently large.  Here we adopt a prescription based on the two-stream approximation \citep{dimatteo+02, kohri+05}:
\begin{eqnarray}
Q^-_{\nu}&=&2\sum_i \frac{(7/8)\sigma T^4}{(3/4)(\tau_{\nu_i}/2+1/\sqrt{3}+1/3\tau_{a,\nu_i})}, \\
u_{\nu}&=&\sum_i \frac{(7/8)aT^4(\tau_{\nu_i}/2+1/\sqrt{3})}{\tau_{\nu_i}/2+1/\sqrt{3}+1/3\tau_{a,\nu_i}},
\end{eqnarray}
where $\tau_{\nu_i}=\tau_{a,\nu_i}+\tau_{s,\nu_i}$ and the neutrino flavor is labeled by subscript $i = (e,\mu,\tau)$.  $\tau_{a,\nu_i}$ and $\tau_{s,\nu_i}$ represent absorptive and scattering optical depths respectively.  The detailed description of these neutrino opacities can be found in \cite{kawanakamineshige07}.

  When the optical depth is so large that the neutrino diffusion timescale in the accretion flow is longer than the accretion timescale, neutrinos are completely trapped in the accretion flow.  In that state, advective cooling dominates: $Q^-\simeq Q^-_{\rm adv}$.  The accretion flow would also be advection-dominated when the mass accretion rate is small enough that the density and temperature of the disk are so low that neutrinos are not efficiently emitted, but is still much larger than the Eddington accretion rate.

We can divide the hyperaccretion flow into several regimes according to the dominant physical processes taking place.  In particular, the cooling processes (advection, optically-thin neutrino emission and optically-thick neutrino emission) and the pressure components (radiation, baryonic gas, and degenerate electrons) vary from one regime to another.  In the rest of this section, we solve the structure of a hyperaccretion flow in its innermost region as a function of the mass accretion rate and examine the dominant physical processes determining the behavior of the accretion flow in the different regimes. 

\subsection{Numerical Results}
We numerically solve the fundamental equations of an hyperaccretion flow, Eqs. (\ref{masscons})-(\ref{hydrostat}), in the innermost region to find the density, temperature, and scale height as functions of the mass accretion rate $\dot{M}$.  In the course of doing so, we also identify the dominant physical mechanisms operating over a wide range of $\dot{M}$ values.  As the relevant ranges of parameters for the central engine of a GRB, we consider $0.01 \lesssim \alpha \lesssim 0.3$, $1\lesssim M_{\rm BH}/M_{\odot} \lesssim 30$, and $2 \lesssim R_{\rm in}/R_g \lesssim 6$.  Our fiducial parameters are $R_{\rm in}=6R_g$, $M_{\rm BH}=3M_{\odot}$ and $\alpha=0.1$.

\begin{figure*}[tbp]
\epsscale{1.8}
\plotone{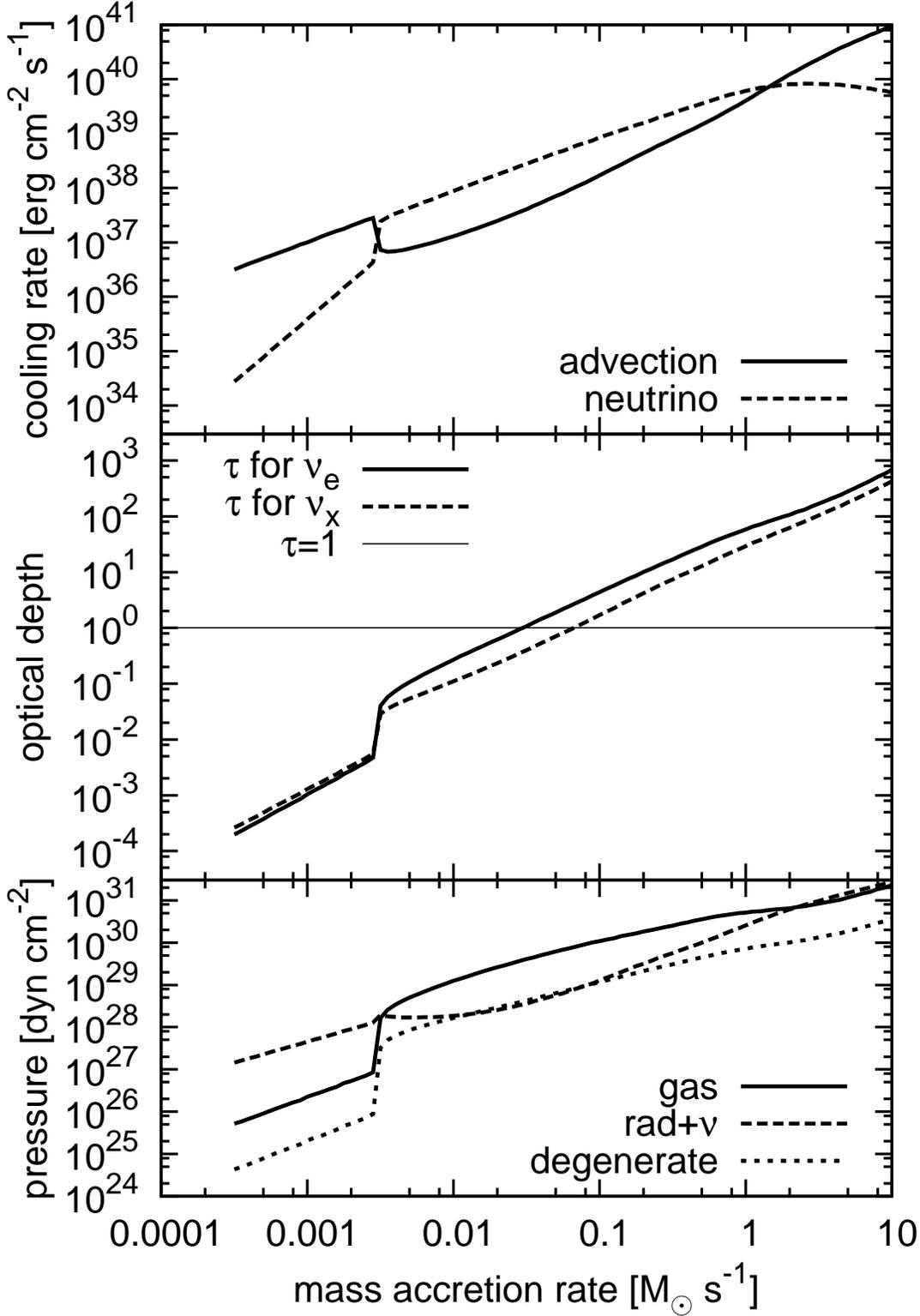}
\caption{Physical properties in the innermost region of a hyperaccretion flow as functions of the mass accretion rate.  Top: cooling rates by advection (solid line) and neutrino emission (dashed line).  Middle: the optical depths for electron neutrinos (thick solid line) and $\mu$/$\tau$ neutrinos (dashed line).  The boundary between optically thin and thick is also shown (thin solid line).  Bottom: the pressure from baryonic gas (solid line), relativistic particles (photons, electron-positron pairs and neutrinos; dashed line) and degenerate electrons (dotted line).}
\label{f1}
\end{figure*}

Figure~\ref{f1} depicts the different cooling rates (neutrino emission and advection), optical depths for different flavors of neutrinos, and pressure components (radiation, baryonic gas, and degenerate electrons) as functions of mass accretion rate.  We can see that neutrino cooling becomes dominant over advective cooling above $\dot{M}\simeq 0.003M_{\odot}{\rm s}^{-1}$.

\begin{figure*}[tbp]
\epsscale{1.8}
\plotone{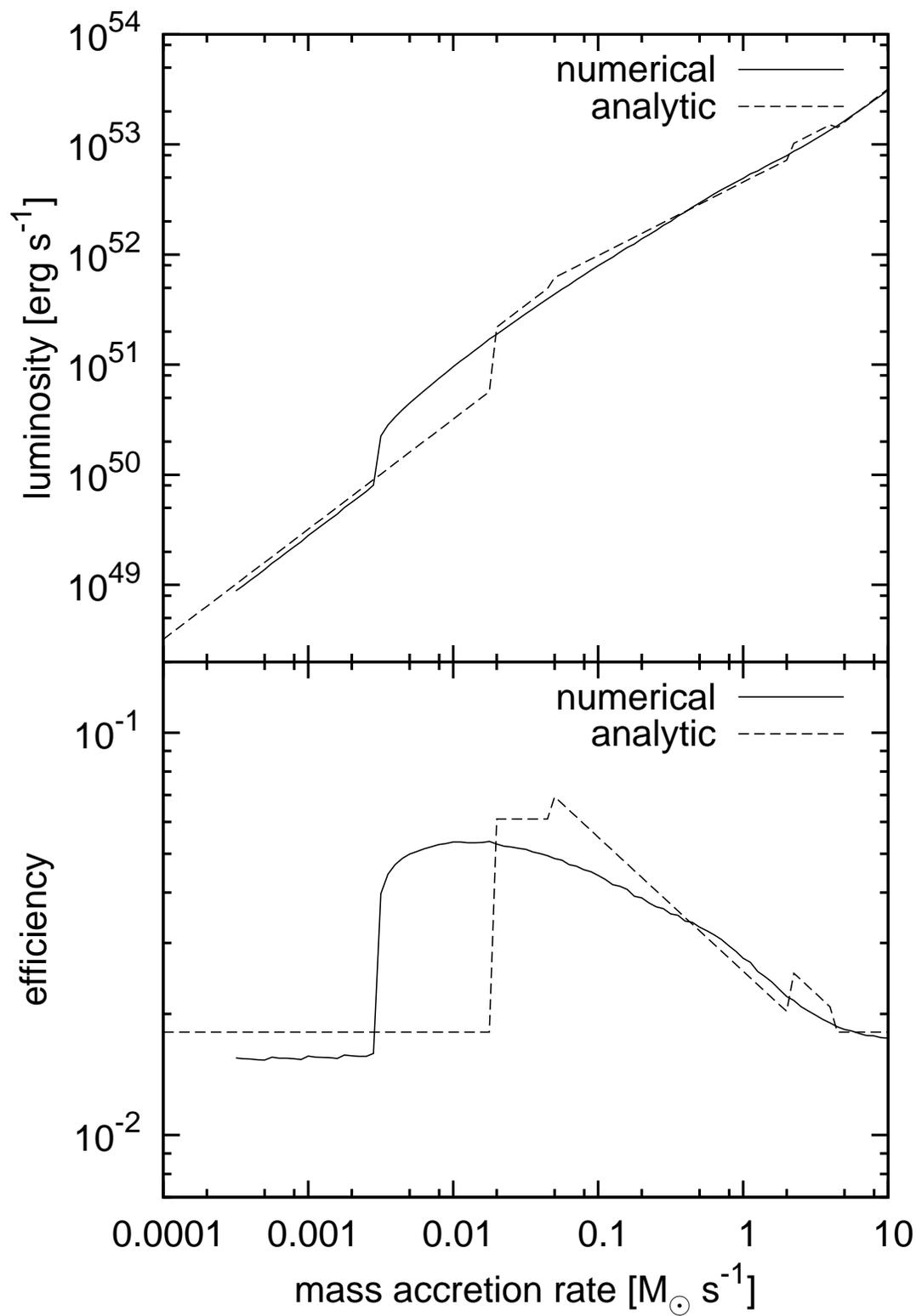}
\caption{Jet luminosities and efficiencies as functions of mass accretion rate calculated numerically (solid line) and analytically (dashed line).  Here $\beta_h$ and $f(a/M_{\rm BH})$ are set to unity.}
\label{f2}
\end{figure*}

Figure~\ref{f2} depicts the jet luminosity $L_{\rm jet}$ and efficiency $\eta_{\rm jet}\equiv L_{\rm jet}/\dot{M}c^2$, estimated from Eq.~(\ref{jetlum}).  Inspection of Figures~\ref{f1} and \ref{f2} reveals a sharp break in the nature of the disk and jet at $\dot{M}\sim 0.003M_{\odot}{\rm s}^{-1}$, the transition from advection to neutrino cooling.  In order to interpret this break, we have constructed an approximate analytical model for hyperaccretion disk structure.  This model can qualitatively reproduce not only the physical properties of the flow, but also the mass accretion rates at which one regime changes to another.

\subsection{The Analytic Model}
\subsubsection{$p\sim p_{\rm rad}$ and advection cooling ($\dot{M}\lesssim 0.018M_{\odot}{\rm s}^{-1}$ for our fiducial parameters)}
When the mass accretion rate is not large enough for the disk to cool efficiently via neutrino emission, the accretion flow is advection-dominated, and the pressure is dominated by photons and relativistic electron/positron pairs.  In this regime, the accretion flow is geometrically thick and the thin-disk approximation is not valid.  Instead, we can estimate the disk pressure using dimensional analysis\footnote{Remarkably, the results are qualitatively similar to those obtained using the thin disk approximation.}.  Because the aspect ratio of the accretion flow $H/R \sim 1$, the density at $R_{\rm in}$ is $\sim \dot{M}/(2\pi R_{\rm in}^2 v_R$).  As the gas does not cool efficiently, the ratio of pressure to density (i.e. the square of the sound speed) is close to the virial temperature $k_B T_{\rm vir} \sim GM_{\rm BH}m_p/R_{\rm in}$.  Using the relation between $v_R$ and the circular orbital speed $v_{\rm orb}$, $v_R/v_{\rm orb} \sim \alpha$, the pressure can be evaluated as:
\begin{eqnarray}
p_{\rm disk}(R_{\rm in})&\sim &\rho c_s^2 \nonumber \\
& \sim & \frac{\dot{M}}{2\pi R_{\rm in}^2 \alpha} \left(\frac{GM_{\rm BH}}{R_{\rm in}} \right)^{1/2} \nonumber \\
&\sim & 5.4\times 10^{27}{\rm dyn}~{\rm cm}^{-2}~\alpha_{0.1}^{-1}(\dot{M}/10^{-3}M_{\odot}{\rm s}^{-1}) \nonumber \\
&&\times (M_{\rm BH}/3M_{\odot})^{-2}(R_{\rm in}/6R_g)^{-5/2}. \label{pres1}
\end{eqnarray}
Using this expression for $p_{\rm disk}$ we estimate the jet luminosity as
\begin{eqnarray}
L_{\rm jet}&\simeq &3.2\times10^{49}{\rm erg}~{\rm s}^{-1}~\beta_h^{-1}\alpha_{0.1}^{-1}(\dot{M}/10^{-3}M_{\odot}{\rm s}^{-1})\nonumber \\
&&\times (R_{\rm in}/6R_g)^{-5/2}f(a/M_{\rm BH}).
\end{eqnarray}
The corresponding energy efficiency of the jet:
\begin{eqnarray}
\eta_{\rm jet}\simeq 0.018\beta_h^{-1}\alpha_{0.1}^{-1}(R_{\rm in}/6R_g)^{-5/2}f(a/M_{\rm BH}) \label{eff1}
\end{eqnarray}
is independent of the mass accretion rate.

The accretion flow is advection-dominated as long as the neutrino emissivity per unit volume is smaller than half the advective cooling rate.  In this regime, neutrino emission is dominated by electron/positron capture, and the neutrino emissivity $q_{\nu}^-$ is given by
\begin{eqnarray}
q_{\nu}^-&\simeq &9.0\times 10^{33}{\rm erg}~{\rm cm}^{-2}~{\rm s}~\rho_{10}T_{11}^6, \label{ecap}
\end{eqnarray}
where $\rho_{10}\equiv \rho/10^{10}{\rm g}~{\rm cm}^{-3}$ and $T_{11}\equiv T/10^{11}{\rm K}$ (Popham, Woosley \& Fryer 1999).  The density and temperature of the gas are:
\begin{eqnarray}
\rho&\sim &\frac{\dot{M}}{2\pi R_{\rm in}^2\alpha v_{\rm orb}} \nonumber \\
&\simeq &3.6\times 10^7{\rm g}~{\rm cm}^{-3}~\alpha_{0.1}^{-1}(\dot{M}/10^{-3}M_{\odot}{\rm s}^{-1}) \nonumber \\
&&\times (M_{\rm BH}/3M_{\odot})^{-2}(R_{\rm in}/6R_g)^{-3/2}, \\
T&\sim &\left(\frac{12p_{\rm disk}}{11a}\right)^{1/4} \nonumber \\
&\simeq& 3.0\times 10^{10}{\rm K}~\alpha_{0.1}^{-1/4}(\dot{M}/10^{-3}M_{\odot}{\rm s}^{-1})^{1/4} \nonumber \\
&&\times (M_{\rm BH}/3M_{\odot})^{-1/2}(R_{\rm in}/6R_g)^{-5/8}.
\end{eqnarray}
Substituting these values into the expression for neutrino emissivity, we obtain:
\begin{eqnarray}
q_{\nu}^-&\simeq 2.3 &\times 10^{28}{\rm erg}~{\rm cm}^{-2}~{\rm s}^{-1}~\alpha_{0.1}^{-5/2}(\dot{M}/10^{-3}M_{\odot}{\rm s}^{-1})^{5/2}\nonumber \\
&&\times (M_{\rm BH}/3M_{\odot})^{-5}(R_{\rm in}/6R_g)^{-21/4}.\label{rhotempadvect}
\end{eqnarray}
On the other hand, the advective cooling rate per unit volume is
\begin{eqnarray}
q_{\rm adv}^-&\sim &Tv_R \rho \left| \frac{ds}{dR} \right| \nonumber \\
&\sim& 3.6\times 10^{30}{\rm erg}~{\rm cm}^{-3}~{\rm s}^{-1}~(\dot{M}/10^{-3}M_{\odot}{\rm s}^{-1}) \nonumber \\
&&\times (M_{\rm BH}/3M_{\odot})^{-3}(R_{\rm in}/6R_g)^{-4}.
\end{eqnarray}
Here we use $s=\gamma p_{\rm disk}/\rho T (\gamma-1)$ where $\gamma=4/3$ is the adiabatic index of relativistic particles.

A transition takes place when $q_{\nu}^-\gtrsim (1/2)q_{\rm adv}^-$, i.e. 
\begin{eqnarray}
\dot{M}/M_{\odot}{\rm s}^{-1}\simeq 0.018\alpha_{0.1}^{5/3}(M_{\rm BH}/3M_{\odot})^{4/3} (R_{\rm in}/6R_g)^{5/6}.
\end{eqnarray}
Above this accretion rate the accretion flow is no longer advection-dominated, as neutrino cooling determines the structure of the flow.  In addition, the dominant pressure source is no longer radiation, but baryonic gas pressure.  In the following, we call this transition mass accretion rate  $\dot{M}_{\rm ign}$ ("ign" stands for ignition of neutrino emission).

\subsubsection{$p\sim p_{\rm gas}$ and optically thin $\nu_e$ cooling ($0.018M_{\odot}{\rm s}^{-1}\lesssim \dot{M}\lesssim 0.045M_{\odot}{\rm s}^{-1}$ for our fiducial parameters)}
In this regime the disk is mainly cooled by neutrino emission via electron/positron capture, while the pressure is dominated by baryonic gas (see \citealp{kawanakamineshige07} and references therein).  In this case, the pressure can be expressed as
\begin{eqnarray}
p_{\rm disk}\simeq \frac{\rho k_B T}{m_p}.
\end{eqnarray}
The neutrino cooling rate per unit surface area by electron/positron capture is given by Eq.~(\ref{ecap}).  Substituting $Q^-=Q_{\nu}^-$ in Eq.~(\ref{energy}), we solve the basic equations and derive the disk properties in this regime:

\begin{eqnarray}
\rho&=&6.3\times 10^{10}{\rm g}~{\rm cm}^{-3}~\alpha_{0.1}^{-13/10}(\dot{M}/0.01M_{\odot}{\rm s}^{-1}) \nonumber \\
&&\times (M_{\rm BH}/3M_{\odot})^{-17/10}(R_{\rm in}/6R_g)^{-51/20}, \\
T&=&3.6\times 10^{10}{\rm K}~\alpha_{0.1}^{1/5}(M_{\rm BH}/3M_{\odot})^{-1/5} \nonumber \\
&&\times (R_{\rm in}/6R_g)^{-3/10}, \\
H&=&3.8\times 10^5{\rm cm}~\alpha_{0.1}^{1/10}(M_{\rm BH}/3M_{\odot})^{9/10} \nonumber \\
&&\times (R_{\rm in}/6R_g)^{27/20}.
\end{eqnarray}
The disk pressure is given by
\begin{eqnarray}
p_{\rm disk}&\simeq &1.9\times 10^{29}{\rm dyn}~{\rm cm}^{-2}~\alpha_{0.1}^{-11/10}(\dot{M}/0.01M_{\odot}{\rm s}^{-1}) \nonumber \\
&&\times (M_{\rm BH}/3M_{\odot})^{-19/10}(R_{\rm in}/6R_g)^{-57/20}.
\end{eqnarray}
Finally we estimate the jet luminosity:
\begin{eqnarray}
L_{\rm jet}&\sim &1.1\times 10^{51}{\rm erg}~{\rm s}^{-1}~\beta_h^{-1}\alpha_{0.1}^{-11/10}(\dot{M}/0.01M_{\odot}{\rm s}^{-1}) \nonumber \\
&&\times (M_{\rm BH}/3M_{\odot})^{1/10}(R_{\rm in}/6R_g)^{-57/20}f(a/M_{\rm BH}).
\end{eqnarray}
The jet luminosity in this regime depends linearly on the mass accretion rate.  The efficiency $\eta_{\rm jet}\equiv L_{\rm jet}/\dot{M}c^2$ is:
\begin{eqnarray}
\eta_{\rm jet}&\sim &0.061\beta_h^{-1}\alpha_{0.1}^{-11/10}(M_{\rm BH}/3M_{\odot})^{1/10} (R_{\rm in}/6R_g)^{-57/20} \nonumber \\
&&\times f(a/M_{\rm BH}),
\end{eqnarray}
and is independent of the mass accretion rate.  The higher jet luminosity and efficiency in this regime is the result of more efficient neutrino cooling, which leads to a geometrically thinner accretion flow.  Thinner disks accrete more slowly.  In the end, the midplane pressure $p_{\rm disk}$ increases, permitting it to confine a stronger magnetic field on the black hole's event horizon.  This stronger magnetic field supports a more powerful jet.

The system is very optically thin to neutrinos at this stage:
\begin{eqnarray}
\tau_{{\rm a},\nu_e}&\simeq & 4.5\times 10^{-7}T_{11}^2\rho_{10}H, \\
\tau_{{\rm s},\nu_i}&\simeq& 2.7\times 10^{-7}T_{11}^2\rho_{10}H, \label{taus}
\end{eqnarray}
where $\tau_{{\rm a},\nu_e}$ and $\tau_{{\rm s},\nu_e}$ are the optical depth due to nucleon absorption and scattering, respectively (see \citealp{dimatteo+02}).  The total optical depth for electron neutrinos $\tau_{\nu_e}=\tau_{{\rm a},\nu_e}+\tau_{{\rm s},\nu_e}$ exceeds unity only when
\begin{eqnarray}
\dot{M}/{M_{\odot}{\rm s}^{-1}}&\gtrsim &0.045\alpha_{0.1}^{4/5}(M_{\rm BH}/3M_{\odot})^{6/5} \nonumber \\
&&\times (R_{\rm in}/6R_g)^{9/5}, \label{trans2}
\end{eqnarray}
and then our expression for $Q^-$ would not be valid.  This mass accretion rate marks the transition to the third regime, which is optically thick to $\nu_e$.  In the following we call this transition $\dot{M}_{\rm thick}$.

\subsubsection{$p\sim p_{\rm gas}$ and optically thick $\nu_e$ cooling ($0.045M_{\odot}{\rm s}^{-1}\lesssim \dot{M}\lesssim 2.2M_{\odot}{\rm s}^{-1}$ for our fiducial parameters)}
In this regime the innermost region of the disk is optically thick to electron neutrinos, and the opacity is dominated by absorption onto nucleons.  While the disk is not so optically thick that the neutrinos are totally trapped in the flow, the cooling rate is:
\begin{eqnarray}
Q^-_{\nu}=2\frac{4(7/2\sigma T^4)}{3\tau_{\nu_e}},
\end{eqnarray}
where $\tau_{\nu}$ is the neutrino optical depth.   

First we consider the case in which the disk is optically thick only to $\nu_e$ and thin for $\nu_{\mu}$ and $\nu_{\tau}$.  With cooling only by electron neutrino emission, the cooling rate is
\begin{eqnarray}
Q^-_{\nu}\approx 7.4\times 10^{34}{\rm erg}~{\rm cm}^{-2}~{\rm s}^{-1}~T^2\rho^{-1}H^{-1},
\end{eqnarray}
which should be equated with the heating rate in the energy balance equation.  Solving the fundamental equations of the disk we derive its density, temperature and scale height:
\begin{eqnarray}
\rho&\approx& 4.1\times 10^{11}{\rm g}~{\rm cm}^{-3}~\alpha_{0.1}^{-1/2}(M_{\rm BH}/3M_{\odot})^{-1/2} \nonumber \\
&&\times (R_{\rm in}/6R_g)^{-3/4}, \\
T&\approx&4.8\times 10^{10}{\rm K}~\alpha_{0.1}^{-1/3}(\dot{M}/0.1M_{\odot}{\rm s}^{-1})^{2/3}(M_{\rm BH}/3M_{\odot})^{-1} \nonumber \\
&&\times (R_{\rm in}/6R_g)^{-3/2}, \\
H&\approx&4.4\times 10^5{\rm cm}~\alpha_{0.1}^{-1/6}(\dot{M}/0.1M_{\odot}{\rm s}^{-1})^{1/3}(M_{\rm BH}/3M_{\odot})^{1/2} \nonumber \\
&&\times (R_{\rm in}/6R_g)^{3/4}.
\end{eqnarray}
The disk pressure in the midplane is
\begin{eqnarray}
p_{\rm disk}&\approx &1.6\times 10^{30}\alpha_{0.1}^{-5/6}(\dot{M}/0.1M_{\odot}{\rm s}^{-1})^{2/3}(M_{\rm BH}/3M_{\odot})^{-3/2} \nonumber \\
&&\times (R_{\rm in}/6R_g)^{-9/4},
\end{eqnarray}
and the jet luminosity is
\begin{eqnarray}
L_{\rm jet}&\sim &9.8\times 10^{51}{\rm erg}~{\rm s}^{-1}~\beta_h^{-1}\alpha_{0.1}^{-5/6}(\dot{M}/0.1M_{\odot}{\rm s}^{-1})^{2/3} \nonumber \\
&&\times (M_{\rm BH}/3M_{\odot})^{1/2}(R_{\rm in}/6R_g)^{-9/8}f(a/M_{\rm BH}).
\end{eqnarray}
The dependence on the mass accretion rate is weaker than in the last regime.  The efficiency is:
\begin{eqnarray}
\eta_{\rm jet}&\sim&0.055\beta_h^{-1}\alpha_{0.1}^{-5/6}(\dot{M}/0.1M_{\odot}{\rm s}^{-1})^{-1/3}(M_{\rm BH}/3M_{\odot})^{1/2} \nonumber \\
&&\times (R_{\rm in}/6R_g)^{-9/4}f(a/M_{\rm BH}).
\end{eqnarray}
In order for the disk to cool by neutrinos efficiently, the neutrino diffusion time $t_{{\rm diff},\nu_e}\equiv H\tau_{\nu_e}/c$ must be shorter than the accretion time $t_{\rm acc}\equiv 1/(\alpha \Omega)(R/H)^2$.  This condition can be expressed as 
\begin{eqnarray}
\dot{M}/{M}_{\odot}{\rm s}^{-1}&\lesssim &2.2\alpha_{0.1}^{5/16}(M_{\rm BH}/3M_{\odot})^{21/16}\nonumber \\
&&\times (R_{\rm in}/6R_g)^{51/32},
\end{eqnarray}
and when the mass accretion rate exceeds this value, electron neutrinos are no longer able to escape the disk.  We call it $\dot{M}_{\rm trap,{\nu_e}}$.  For the other kinds of neutrinos ($\nu_{\mu}$ and $\nu_{\tau}$), the diffusion time is still shorter than the accretion time, so they can cool the disk efficiently.  This is the fourth regime, which we discuss in the next subsection.


\subsubsection{$p\sim p_{\rm gas}$, and optically thick $\nu_x$ cooling ($2.2M_{\odot}{\rm s}^{-1}\lesssim \dot{M}\lesssim 4.1M_{\odot}{\rm s}^{-1}$ for our fiducial parameters)}
Even when the mass accretion rate is so large that $\nu_e$s are completely trapped in the disk, $\nu_{\mu}$s and $\nu_{\tau}$s emitted via the electron-positron pair annihilation process may escape the disk so long as their diffusion time is shorter than the accretion time.  For those neutrinos, the opacity is dominated by scattering on nucleons (Eq.~\ref{taus}).  Then the cooling rate is
\begin{eqnarray}
Q^-_{\nu}&\simeq &2\times \left(2\frac{4(7/2)\sigma T^4}{3\tau_{\nu_x}}\right) \nonumber \\
&\simeq &2.0\times 10^{35}{\rm erg}~{\rm cm}^{-2}~{\rm s}^{-1}~T_{11}^2 \rho_{10}^{-1}H,
\end{eqnarray}
where the factor 2 in the first line represents the number of flavors of neutrinos that we are interested in ($\nu_{\mu}$/$\bar{\nu}_{\mu}$ and $\nu_{\tau}$/$\bar{\nu}_{\tau}$).  Substituting this $Q^-$ into Eq. (\ref{energy}), we derive the disk properties:
\begin{eqnarray}
\rho&\simeq&9.3\times 10^{11}{\rm g}~{\rm cm}^{-3}~\alpha_{0.1}^{-1/2}(M_{\rm BH}/3M_{\odot})^{-1/2} \nonumber \\
&&\times (R_{\rm in}/6R_g)^{-3/4}, \\
T&\simeq&1.3\times 10^{11}{\rm K}~\alpha_{0.1}^{-1/3}(\dot{M}/1M_{\odot}{\rm s}^{-1})^{2/3}(M_{\rm BH}/3M_{\odot})^{-1} \nonumber \\
&&\times (R_{\rm in}/6R_g)^{-3/2}, \\
H&\simeq&7.3\times 10^5{\rm cm}~\alpha_{0.1}^{-1/6}(\dot{M}/1M_{\odot}{\rm s}^{-1})^{1/3}(M_{\rm BH}/3M_{\odot})^{1/2} \nonumber \\
&&\times (R_{\rm in}/6R_g)^{3/4}.
\end{eqnarray}
The disk pressure is then
\begin{eqnarray}
p_{\rm disk}&\simeq&9.9\times 10^{30}{\rm dyn}~{\rm cm}^{-2}~\alpha_{0.1}^{-5/6}(\dot{M}/1M_{\odot}{\rm s}^{-1})^{2/3} \nonumber \\
&&\times (M_{\rm BH}/3M_{\odot})^{-3/2}(R_{\rm in}/6R_g)^{-9/4},
\end{eqnarray}
and the jet luminosity is
\begin{eqnarray}
L_{\rm jet}&\sim &6.0\times 10^{52}{\rm erg}~{\rm s}^{-1}~\beta_h^{-1}\alpha_{0.1}^{-5/6}(\dot{M}/1M_{\odot}{\rm s}^{-1})^{2/3} \nonumber \\
&&\times (M_{\rm BH}/3M_{\odot})^{1/2}(R_{\rm in}/6R_g)^{-9/4}f(a/M_{\rm BH}).
\end{eqnarray}
The luminosity now depends on the mass accretion rate in the same way as in the last regime.  The efficiency is
\begin{eqnarray}
\eta_{\rm jet}&\sim &0.033\beta_h^{-1}\alpha_{0.1}^{-5/6}(\dot{M}/1M_{\odot}{\rm s}^{-1})^{-1/3}(M_{\rm BH}/3M_{\odot})^{1/2} \nonumber \\
&&\times (R_{\rm in}/6R_g)^{-9/4}f(a/M_{\rm BH}).
\end{eqnarray}
The timescale for $\mu$/$\tau$ neutrinos to escape the disk diffusively is
\begin{eqnarray}
t_{{\rm diff},\nu_x}&\simeq&\frac{H\tau_{{\rm s},\nu_x}}{c} \nonumber \\
&\simeq&5.2\times 10^{-4}{\rm s}~\alpha_{0.1}^{-3/2}(\dot{M}/1M_{\odot}{\rm s}^{-1})^2(M_{\rm BH}/3M_{\odot})^{-3/2} \nonumber \\
&&\times (R_{\rm in}/6R_g)^{-9/4},
\end{eqnarray}
and the accretion timescale is
\begin{eqnarray}
t_{\rm acc}&\simeq&\frac{1}{\alpha \Omega}\left( \frac{R}{H} \right)^2 \nonumber \\
&\simeq&3.9\times 10^{-2}{\rm s}~\alpha_{0.1}^{-2/3}(\dot{M}/1M_{\odot}{\rm s}^{-1})^{-2/3}(M_{\rm BH}/3M_{\odot})^2 \nonumber \\
&&\times (R_{\rm in}/6R_g)^2.
\end{eqnarray}
Therefore, $\mu$/$\tau$ neutrinos can escape an accretion flow without being trapped as long as
\begin{eqnarray}
\dot{M}/M_{\odot}{\rm s}^{-1}&\lesssim &4.1\alpha_{0.1}^{5/16}(M_{\rm BH}/3M_{\odot})^{21/16}\nonumber \\
&&\times (R_{\rm in}/6R_g)^{51/32}.
\end{eqnarray}
At this mass accretion rate, there is a transition to the fifth and last regime, in which the accretion flow is once more advection-dominated.

\subsubsection{$p\sim p_{\rm rad}+p_{e^{\pm}}+\Sigma p_{\nu_i}$, $Q^-\sim Q^-_{\rm adv}$ ($4.1M_{\odot}{\rm s}^{-1}\lesssim \dot{M}$ for our fiducial parameters)} 
When the mass accretion rate is so large that the all kinds of neutrinos are trapped, the accretion flow cannot cool and the flow is advection-dominated.  In this case the pressure would be dominated by several kinds of relativistic particles, photons, relativistic electron-positron pairs, and all three flavors of neutrinos.   We can estimate the disk pressure in a way similar to Sec.~2.2.1 to find
\begin{eqnarray}
p_{\rm disk}&\simeq &2.7\times 10^{31}{\rm dyn}~{\rm cm}^{-2}~\alpha_{0.1}^{-1}(\dot{M}/10M_{\odot}{\rm s}^{-1})\nonumber \\
&&\times (M_{\rm BH}/3M_{\odot})^{-2}(R_{\rm in}/6R_g)^{-5/2}.
\end{eqnarray}
The jet luminosity is then
\begin{eqnarray}
L_{\rm jet}&\simeq &3.2\times 10^{53}{\rm erg}~{\rm s}^{-1}~\beta_h^{-1} \alpha_{0.1}^{-1}(\dot{M}/10M_{\odot}{\rm s}^{-1})\nonumber \\
&&\times (R_{\rm in}/6R_g)^{-5/2}f(a/M_{\rm BH}).
\end{eqnarray}
As a result, the efficiency is the same as in Equation~(\ref{eff1}),
\begin{eqnarray}
\eta_{\rm jet}\simeq 0.018\beta_h^{-1}\alpha_{0.1}^{-1}(R_{\rm in}/6R_g)^{-5/2} f(a/M_{\rm BH}).
\end{eqnarray}

\begin{figure}[h]
\epsscale{0.75}
\plotone{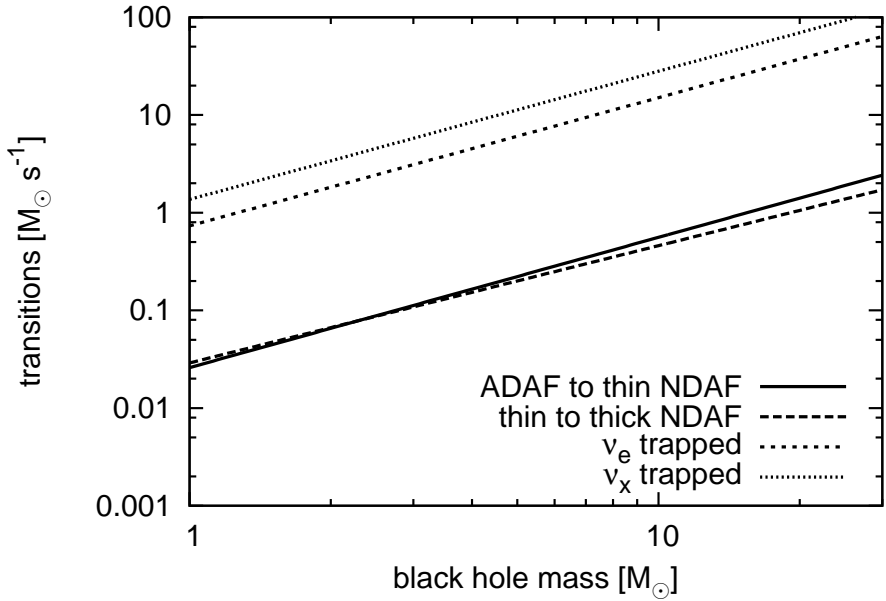}
\caption{The four transition mass accretion rates as functions of the black hole mass.  Here we fix the other parameters as $\alpha=0.3$ and $R_{\rm in}=6R_g$.  Note that, for canonical values of these parameters (i.e. $\alpha=0.1$ and $R_{\rm in}=6R_g$), there is no change of the order of transition points in this black hole mass range.}
\label{f3}
\end{figure}

\subsection{The Order of the Transition Points}
\begin{figure}[h]
\epsscale{0.75}
\plotone{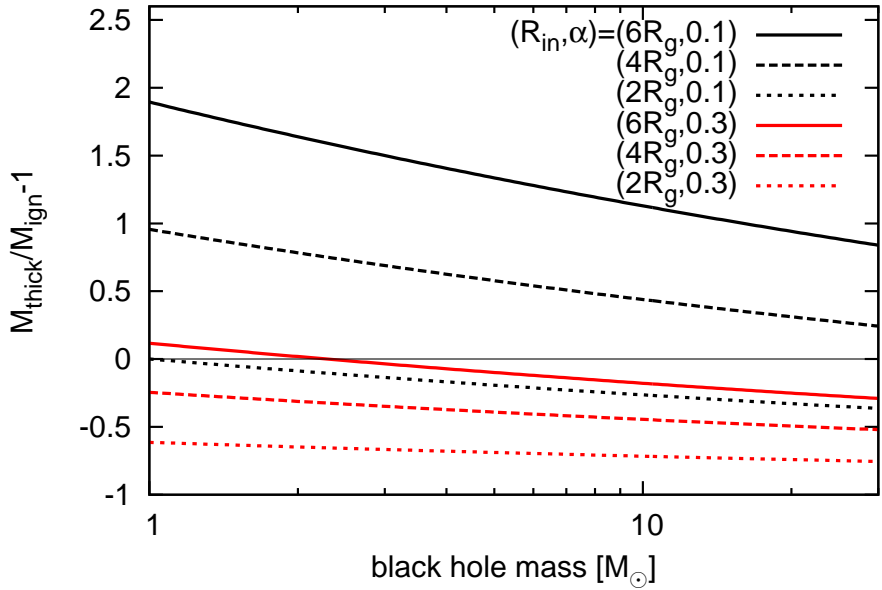}
\caption{$\dot{M}_{\rm thick}/\dot{M}_{\rm ign}-1$ as a function of the black hole mass.  Black and red lines represent the case with $\alpha=0.1$ and $0.3$, respectively, and solid, dashed and dotted lines represent the case with $R_{\rm in}=6R_g$, $4R_g$ and $2R_g$, respectively.  The horizontal thin black line represents $\dot{M}_{\rm thick}/\dot{M}_{\rm ign}=1$.}
\label{f4}
\end{figure}

\begin{figure}[h]
\epsscale{0.75}
\plotone{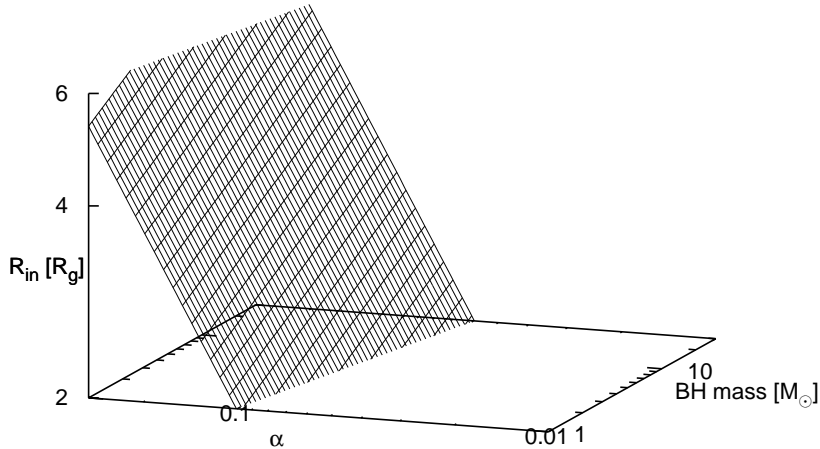}
\caption{The optically thin neutrino cooling regime exists above (and to the right) of the marked plane in the parameter phase space of ($\alpha$, $M_{\rm BH}$, $R_{\rm in}$).}
\label{f5}
\end{figure}

In the discussion above, we have picked out the four mass accretion rates that divide the five states of an hyperaccretion flow.  However, the sequence we have presented does not always apply.  These mass accretion rates are functions of $\alpha$, $M_{\rm BH}$, and $R_{\rm in}$, and in some regions of parameter space, the transition from advection-dominated flow to neutrino-cooled flow ($\dot{M}_{\rm ign}$) takes place at a higher accretion rate than the transition to optically thick flow (with respect to $\nu_e$).  This means that the second regime of optically thick neutrino cooling does not exist for these parameters.  For example, we show the transition mass accretion rates as functions of the black hole mass (while the other parameters are fixed at $\alpha=0.3$ and $R_{\rm in}=6R_g$) in Figure~\ref{f3}.  One can see that there is a critical black hole mass at which the first transition and the second transition change places, and also that for this choice of parameters the mass accretion rate range of the optically-thin NDAF regime is quite narrow, if it exists at all.  In Figure~\ref{f4}
we show $\dot{M}_{\rm thick}/\dot{M}_{\rm ign}-1$ for various parameter sets as functions of the black hole mass.  For $(R_{\rm in},\alpha)=(6R_g,0.1)$ and $(4R_g,0.1)$, the curve is well above $\dot{M}_{\rm thick}/\dot{M}_{\rm ign}-1$, and the optically-thin NDAF regime exists for each of those parameter sets.  The general condition for such a change in the order of the transition points is:
\begin{eqnarray}
\alpha^{13/15}(M_{\rm BH}/M_{\odot})^{2/15}(R/R_g)^{-29/30} \gtrsim 0.07. \label{change}
\end{eqnarray}
When this condition is satisfied, advection-dominated flow transforms directly to optically thick neutrino cooling at (\ref{trans2}).  Figure~\ref{f5}
displays the limited range of parameters satisfying this condition.  In other words, except in the case of extreme values of the parameters, the five states for hyperaccretion flows are ordered as we have described them.

\section{Results and Discussion}

\subsection{Parameter Dependence of the Jet Luminosity}
\begin{figure}
\epsscale{1.5}
\plotone{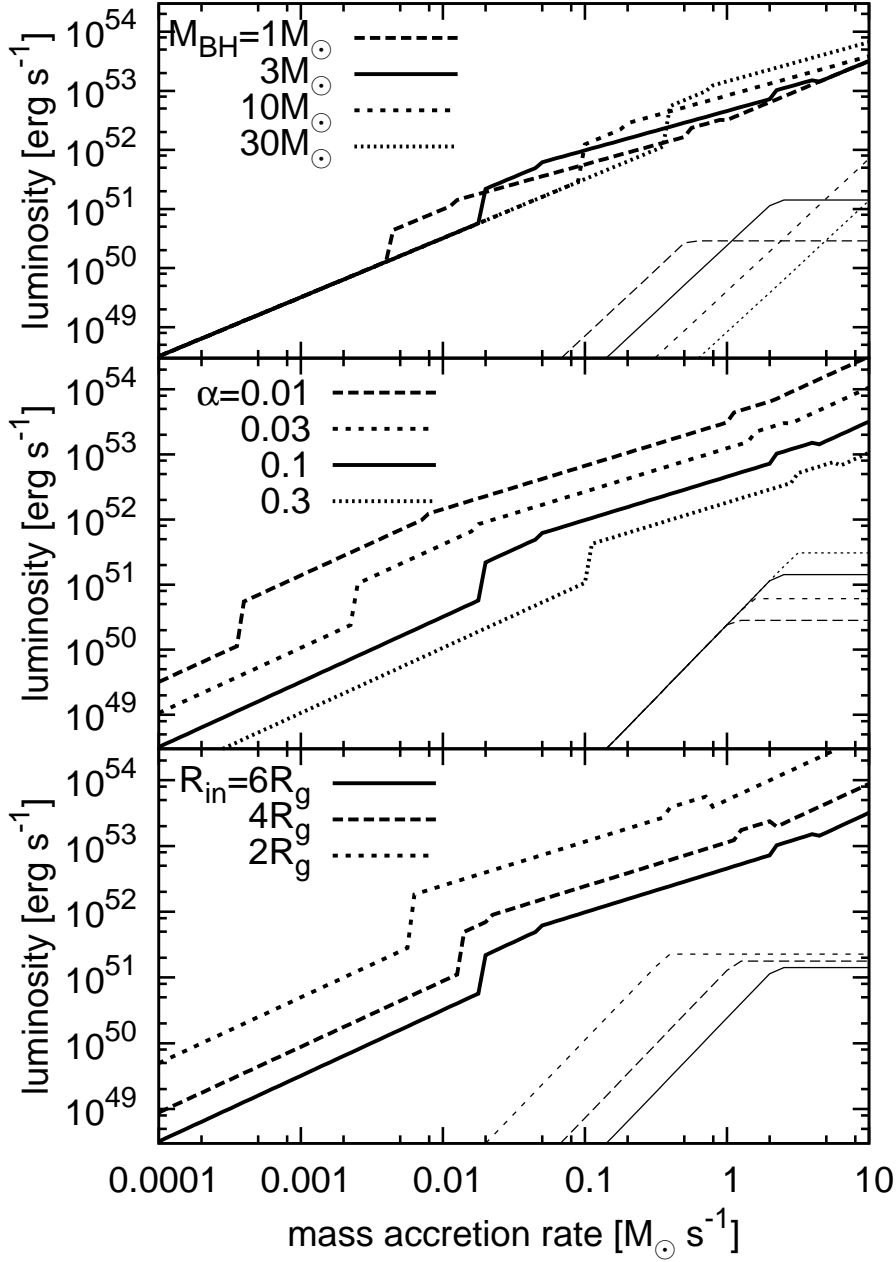}
\caption{Jet luminosity expected from the BZ mechanism (thick lines) and neutrino pair annihilation (thin lines).  Their dependence on the central black hole mass ({\it top}), on $\alpha$ ({\it middle}) and on the inner radius of an accretion flow ({\it bottom}) are shown.  The model parameters are $\alpha=0.1$, $M_{\rm BH}=3M_{\odot}$ and $R_{\rm in}=6R_g$ unless otherwise stated.  Here $\beta_h$ and $f(a/M_{\rm BH})$ are set to unity.}
\label{f6}
\end{figure}

From the analytic model presented above, we can estimate the jet luminosity and efficiency expected from a hyperaccreton flow as functions of mass accretion rate for various choices of parameters ($\alpha$, $M_{\rm BH}$, $R_{\rm in}$).  Figure~\ref{f6} shows how the jet luminosity based on the BZ scenario (thick lines) depends on the mass accretion rate.  The different lines show the results for different parameters.  We can see that with all parameter sets, the jet luminosity has a significant discontinuity at $\dot{M}_{\rm ign}$.  This jump, as explained already, arises from the transition between an advection-dominated accretion flow, in which neutrino cooling is not efficient, and a neutrino-dominated accretion flow.
In the top panel, we can see that for larger black hole mass the mass accretion rate at which the transition from the advection-dominated flow to the neutrino-cooled flow ($\dot{M}_{\rm ign}$) is larger.  The value of $\dot{M}_{\rm ign}$ rises from $\simeq 5 \times 10^{-3} M_{\odot}$~s$^{-1}$ at $M_{\rm BH}=1M_{\odot}$ to $\simeq 0.3 M_{\odot}$~s$^{-1}$ at $M_{\rm BH} = 30M_{\odot}$.  This is because the temperature of the accretion flow is lower for larger black hole mass when the radius (in units of the gravitational radius) and the mass accretion rate are fixed.  Therefore, a larger mass accretion rate is needed for an accretion flow around a larger black hole to cool efficiently via neutrino emission.

A similar explanation also applies to the bottom panel, which shows the dependence of the jet luminosity on $R_{\rm in}$.  Here, for smaller $R_{\rm in}$ the temperature of the accretion flow is higher, which results in a smaller $\dot{M}_{\rm ign}$ and larger luminosity.  Over the range of plausible values of $R_{\rm in}$, the variation of $\dot{M}_{\rm ign}$ is much smaller than for the dependence on $M_{BH}$: it changes by only a factor $\sim 4$.

The middle panel shows that the jet luminosity depends quite strongly on the stress/pressure ratio in the sense that for larger $\alpha$, the luminosity is smaller; equivalently, $\dot{M}_{\rm ign}$ increases with increasing $\alpha$.  Over a span of a factor of 30 in $\alpha$, from $\alpha = 0.01$ to $\alpha=0.3$, $\dot{M}_{\rm ign}$ increases by a factor of 300.
To interpret this behavior, we note that the inflow rate is proportional to $\alpha$ (i.e. the accretion timescale is proportional to $\alpha^{-1}$).  Therefore, when $\alpha$ is larger in the radiation-dominated advective regime, the midplane pressure is smaller (eqn.~\ref{pres1}).  The temperature is also lower (eqn.~\ref{rhotempadvect}).  These two trends together diminish the neutrino cooling rate, forcing the transition accretion rate higher.  They are reinforced by the scalings on the opposite side of the transition, in which $\rho \propto \alpha^{-13/10}$, while the temperature depends much more weakly on $\alpha$ (only $\propto \alpha^{1/5}$).  Thus, the jet luminosity is strongly reduced by a larger stress/pressure ratio.

\begin{figure}
\epsscale{1.7}
\plotone{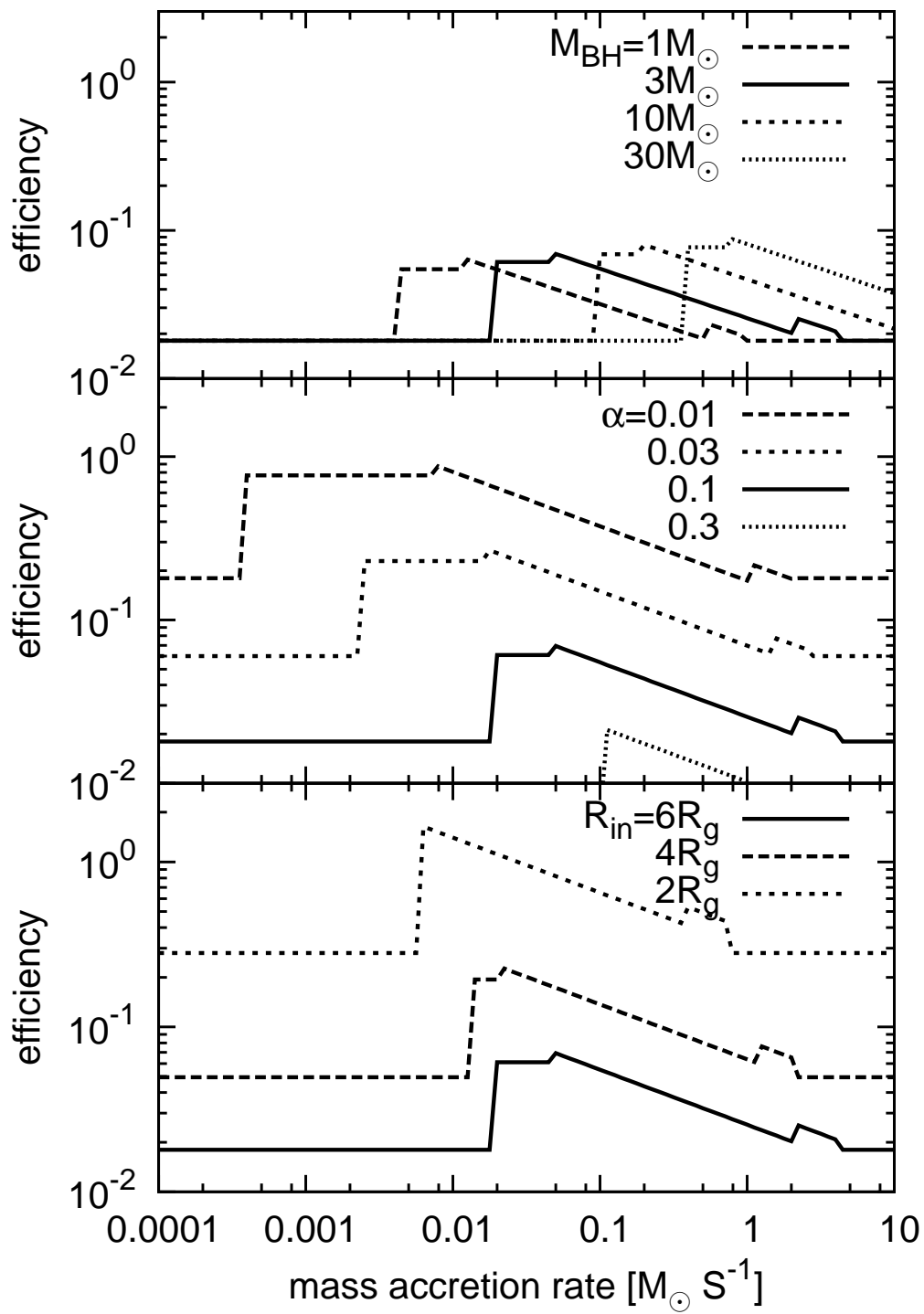}
\caption{Jet efficiency expected from the BZ mechanism.  The parameters are the same as those in Fig. ~\ref{f5}.}
\label{f7}
\end{figure}

Figure~\ref{f7} similarly shows how the jet efficiency depends on our three free parameters.  As we have previously emphasized, the principal dependence is on accretion regime: the jet efficiency is greatest when the flow is in the optically thin neutrino-cooling state.  Although this state occurs over different ranges of accretion rate for different black hole masses, the efficiency achieved there varies rather little with $M_{\rm BH}$.   On the other hand, even after allowing for the different accretion rate ranges in which the flow is neutrino-cooled, it is quite sensitive to both $\alpha$ and $R_{\rm in}/R_g$.  In that favored regime, $\eta_{\rm jet}$ is roughly $\propto \alpha^{-1}$.  Still more dramatically, a factor of 3 decrease in $R_{\rm in}/R_g$ leads to a factor of 30 increase in $\eta_{\rm jet}$ at the respective peaks of jet efficiency in the optically-thin neutrino-cooled state---and this comparison is taken at fixed $f(a/M_{\rm BH})$, whereas realistically one would expect smaller $R_{\rm in}/R_g$ to be associated with more rapid spin and therefore a significant increase in $f(a/M_{\rm BH})$.  { Note that in a small region of the lower panel $\eta_{\rm jet}>1$.  While possible, in principle, in a BZ process, our Newtonian approximation is not suitable to explore this regime.}

\subsection{BZ versus Neutrino Annihilation Jets}
Neutrino-antineutrino annihilation into electrons and positrons (hereafter ``neutrino pair annihilation") above the accretion flow \citep{eichler+89} is another mechanism proposed for the energy source of relativistic jets that power GRBs.  It is worth comparing the luminosity and efficiency of such neutrino pair annihilation driven jets with the BZ driven jets expected from our NDAF model.

In estimating the energy deposition rate due to neutrino pair annihilation above a hyperaccretion flow, we adopt a simplified formulation: the accretion flow is assumed to be geometrically thin (i.e. neutrinos are emitted from the equatorial plane), and we neglect the bending of neutrino paths and the gravitational redshift.  The energy deposition rate per unit volume above the accretion flow can then be calculated from
\begin{eqnarray}
l_{\nu\bar{\nu}}^+(\nu_i \bar{\nu}_i)&=&A_1\sum_k \frac{\Delta L_{\nu_i}^k}{d_k^2}\sum_{k^{\prime}}\frac{\Delta L_{\bar{\nu}_i}^{k^{\prime}}}{d_{k^{\prime}}^2}\left[\langle\epsilon\rangle_{\nu_i}+\langle\epsilon\rangle_{\bar{\nu}_i}\right] (1-\cos \theta)^2 \nonumber \\
&&+A_2\sum_k \frac{\Delta L_{\nu_i}^k}{d_k^2}\sum_{k^{\prime}}\frac{\Delta L_{\bar{\nu}_i}^{k^{\prime}}}{d_{k^{\prime}}^2}\frac{\langle\epsilon\rangle_{\nu_i}+\langle\epsilon\rangle_{\bar{\nu}_i}}{\langle\epsilon\rangle_{\nu_i}\langle\epsilon\rangle_{\bar{\nu}_i}}(1-\cos \theta), \nonumber \\
\end{eqnarray}
where $A_1=\sigma_0(C_1+C_2)_{\nu_i\bar{\nu}_i}/(12\pi^2 c^5 m_e^2)\approx 1.7\times 10^{-44}{\rm cm}~{\rm ergs}^{-2}~{\rm s}^{-1}$ and $A_2=\sigma_0(C_3)_{\nu_i\bar{\nu}_i}/(4\pi^2 c)\approx 1.6\times 10^{-56}{\rm cm}~{\rm s}^{-1}$ are the neutrino cross section constants for electron neutrinos (Ruffert et al. 1997).  Here $\Delta L_{\nu_i}^k$ and $\langle \epsilon \rangle_{\nu_i}$ are the neutrino luminosity coming from the $k$-th cell in the accretion flow and the mean neutrino energy, respectively.  For the neutrino emissivity distribution in the accretion flow, we adopt a simplified model: neutrinos and antineutrinos are emitted with a Fermi-Dirac distribution function with a zero-chemical potential and a temperature
\begin{eqnarray}
T_{\rm eff}(\dot{M},R)&=&T_{\rm eff}^{\rm st}(\dot{M}_{\rm ign},R) \nonumber \\
&&\times\left\{
\begin{array}{ll}
0 & \dot{M}<\dot{M}_{\rm ign} \\
(\dot{M}/\dot{M}_{\rm ign})^{1/4} & \dot{M}_{\rm ign}<\dot{M} \\
(\dot{M}_{\rm trap,{\nu_e}}/\dot{M}_{\rm ign})^{1/4} & \dot{M}>\dot{M}_{\rm trap,{\nu_e}}. \\
\end{array} \right.
\end{eqnarray}
where $T_{\rm eff}^{\rm st}(\dot{M},R)=(Q^+/\sigma)^{1/4}$.  According to Zalamea \& Beloborodov (2011), such a simplification does not change the results significantly.  Integrating $l^+_{\nu\bar{\nu}}$ over the volume for $\theta\lesssim \pi/4$, where $\theta$ is the angle from the rotation axis and $R=0-30R_g$, we obtain the total energy deposition rate via neutrino pair annihilation above the accretion flow.

We contrast the jet luminosity expected from the neutrino pair annihilation scenario with the jet luminosity generated by the BZ mechanism in Fig.~\ref{f6} (thin lines).
In all but the highest accretion rate state, the one in which all three neutrino flavors are very optically thick, we see that the BZ jet luminosity (for our fiducial parameters) is many orders of magnitude greater than the jet luminosity that would be produced by neutrinos.  Even in that state, the neutrino luminosity is generally two orders of magnitude smaller than the BZ jet luminosity.  In the accretion rate range most favorably to the BZ mechanism, the ratio is at least another order of magnitude larger.  Thus, even if $f(a/M_{\rm BH})$ is as small as $\sim 10^{-2}$, the BZ mechanism dominates at all accretion rates; in its favored range, it can be as small as $\sim 10^{-3}$ and still generate a larger luminosity than the neutrino mechanism.

{When comparing the jet peak luminosity with those observed in typical GRBs, $\sim 10^{50}{\rm erg}~{\rm s}^{-1}$, after beaming corrections \citep{ghirlanda+06} we find that with our fiducial parameters those require accretion rate larger than $0.003-0.01M_{\odot}{\rm s}^{-1}$.  The drop by a factor of $\sim 5$ in the efficiency at the transition $\dot{M}_{\rm ign}$ may be a mechanism that strongly modulates the jet power, leading to the observed variability in GRBs.}


\section{Conclusion and Summary}
We have examined the Blandford-Znajek jet luminosity and efficiency expected from hyperaccreting black holes, investigating the possibility that such a jet can account for observed GRBs.  Following \cite{krolikpiran11, krolikpiran12}, we estimated the Poynting luminosity using a simple dimensional argument: we assumed that the intensity of the magnetic field responsible for launching the jet is linked to the pressure in the innermost region of the accretion flow.  By using an NDAF solution at the innermost region of the accretion flow, we have demonstrated that the luminosity of this Poynting-dominated jet can easily exceed the energy deposition rate expected from neutrino pair annihilation above the accretion flow.  This means that MHD processes around the black hole horizon are a plausible mechanism for the formation of the relativistic jets required from the observations of GRBs.  In addition, we have derived the jet luminosity as a function of mass accretion rate and have shown that it has a step function-like behavior at a mass accretion rate corresponding to the transition between an advection-dominated accretion flow and a neutrino-dominated accretion flow.  This means that the jet power suddenly drops when the mass accretion rate decreases and crosses $\dot{M}_{\rm ign}$.  In the context of GRBs, this jump may correspond to switching on and off the activity needed for internal shocks.  It may also explain the steep decay ($\alpha \simeq 3-5; \alpha \equiv |d \log F_{\nu}/d \log t|$) often observed after the prompt emission \citep{tagliaferri+05, nousek+06, obrien+06}.  We have also shown that the MHD jet formation process is most efficient when the accretion flow efficiently cools via optically-thin neutrino emission.  In this regime, the jet luminosity is proportional to the mass accretion rate $\dot{M}$, i.e., the efficiency does not depend on $\dot{M}$.

We stress that the BZ jet luminosity is larger than the energy deposition rate via neutrino pair annihilation whenever $f(a/M_{\rm BH})\gtrsim 0.01$.  In the optimal accretion range of accretion rates, this lower bound is relaxed by at least another order of magnitude.  Thus, for the BZ mechanism to dominate the neutrion mechanism, the central black hole must spin at least moderately fast, and the magnetic field geometry must be reasonably well-organized in the poloidal sense.  However, as this lower bound on $f(a/M_{\rm BH})$ shows, these thresholds are far from extreme.

\acknowledgements{
This work is supported by an Advanced ERC grant (NK and TP) and NSF Grant AST-0908336 (JHK).  The numerical calculations were carried out on SR16000 at Yukawa Institute for Theoretical Physics at Kyoto University.}

\end{document}